\documentstyle[twocolumn,aps]{revtex}
\begin{document}
\noindent{\bf Comment on ``$T$-dependence of
the magnetic penetration depth in
unconventional superconductors at low
temperatures: Can it be linear?"} 
Recently, Schopohl and Dolgov (SD)\cite{SD} raised the possibility that a
$d$-wave superconductor might be thermodynamically unstable in the limit
$T\rightarrow 0$.  Within the framework of linear response theory, they
calculated the entropy of a $d$-wave superconductor due to the
presence of a current,
\begin{eqnarray}
S(T)=-{\partial F\over \partial T}=-\sum_{\bf q} {\partial
\over \partial T}\left[{1\over \lambda^2 ({\bf q},T) }\right]{|B({\bf
q};T)|^2\over 8\pi q^2},
\end{eqnarray}
where $c/4\pi \lambda^2({\bf q},T)$ is the (linear) static current
response function, and $\bf B$ the magnetic field in the sample.  They
argued that the Nernst theorem $\lim_{T\rightarrow 0} ~S\rightarrow 0$
would then require $\partial \lambda ({\bf q},T)/\partial T=0$, whereas
the penetration depth $\lambda(T)$ of a $d$-wave superconductor is known
to be linear in temperature, apparently yielding $S(T\rightarrow 0)=$
const.  While there are several effects which can lead to a nonlinear
$T$-dependence in a  superconductor with line nodes at  low temperatures,
including impurities\cite{impurities} and nonlocal effects,\cite{PHW,KL}
SD pose an intriguing question of principle: Suppose one considers a pure
2D system in a special geometry where the nonlocal effects such as that 
discussed in Ref. \cite{KL} vanish. In such a case it appears as if  
 $\delta\lambda \sim T$.  A constant
entropy at $T\rightarrow 0$ would therefore result, implying the 
breakdown of BCS
theory for a pure $d$-wave state, an unsettling prospect.

One answer to the question posed by SD has been provided by Volovik, 
\cite{Volovik} who
points out that the linear response calculation of SD is not complete,
since any physical experiment will be performed with a finite current. 
He argues that  therefore, when considering Nernst's theorem, the limit
$T\rightarrow 0$ should be taken first.  Based on a semiclassical 
analysis of the resulting nonlinear effects, he shows that the free
energy  varies as
$const+T^2$ rather than $T$ at low temperatures, leading indeed to zero 
residual entropy as required for stability.

While we agree with Volovik's point that nonlinear effects always prevent 
an instability, we believe BCS theory is even more robust than implied by
his argument, and that the answer to the stability question should not
rely sensitively on the order of the limits taken.  
As one lowers the
temperature, there are several ways in which the $d$-wave system can
escape the SD ``paradox", of which the Volovik scenario is only one.
A more general
analysis for a given particular system {\it with or without a surface} 
should include
consideration of both nonlinear and nonlocal effects, as performed
recently for the vortex state.\cite{Aminetal} 
Thus in a pure $d$-wave superconductor, there are three relevant energy
scales: $T$, $E_{nonloc}$ and $E_{nonlin}$.  The magnitude of the last
two energies depend on the particular spatial variations of the currents
in the system (they are not independent in the vortex lattice case).
In a Meissner geometry with {\it generic} surface,
$E_{nonloc}\simeq{\bf v}_F\cdot{\bf q}\simeq
\Delta_0\xi_0/\lambda_0$, where $\Delta_0$ is the gap maximum, and $\xi_0$
and $\lambda_0$ are the zero-temperature coherence length and penetration
depth, respectively.  $E_{nonloc}$ will be larger than the typical
nonlinear scale $E_{nonlin}\simeq {\bf v}_s\cdot {\bf k}_F$, where ${\bf
v}_s$ is the superfluid velocity, if the applied current is sufficiently
small.  In such a case, the response function entering (1) may easily be
shown to be $\lambda^{-2}({\bf q},T)\sim const+T^3$ for  
$T\ll q v_F$ (only after integrating over all Fourier 
components does one obtain the
asymptotic nonlocal result $\delta \lambda \sim T^2$ of Ref.
\cite{KL}).  Thus the ``paradox" is trivially avoided even in linear
response $E_{nonlin}\rightarrow 0$.

As discussed by SD, however, in a special situation with ${\bf H}\perp
{\hat c } \parallel {\hat n}$ in a half-space geometry, it would seem at
first that there are no nonlocal effects, since 
for a strictly 2D order parameter confined to the $ab$ plane, ${\bf
q}\cdot {\bf v}_F =0$.  In this case, however, 
the magnetic field is not screened ($\lambda\rightarrow\infty$)
\cite{GK}. For a quasi-2D system, on the other hand, 
${\bf q}\cdot {\bf v}_F=qv^c_F\neq 0$, and thus nonlocal effects always exist.
The difference $\xi_{{\bf k}+{\bf q}/2}-\xi_{{\bf k}-{\bf q}/2}$ 
($\xi_{\bf k}=k^2/2m-\mu$ is the quasiparticle energy in the normal state)
 which cuts off the nodal singularity in the full response
 also  contains a further term $q^2/2m$ such that in the SD special case the 
energy scale  $E_{nonloc}\simeq \mbox{max}
[\Delta_0\xi_{0c}/\lambda_0, 1/(2m\lambda_0^2)]$, where $\xi_{0c}$
is the c-axis coherence length.

Therefore there appears to be no geometry in which nature does not find a 
way, at
sufficiently low temperature, to force the penetration depth, and hence
the free energy, away from a linear $T$ variation, thus avoiding the SD
``paradox".  A $d$-wave system may be stable down to $T=0$.
\vskip .2cm
\vskip .1cm
{\small P.J. Hirschfeld$^{a,b}$, M. -R. Li$^{b}$,
and P. W\"olfle$^b$
\vskip .08cm
\indent
$^a$ Dept. of Physics, U. Florida, Gainesville,
FL 32611 USA
\vskip .01cm
\indent $^b$ Institut f\"ur Theorie der
Kondensierten Materie, Universit\"at Karlsruhe,
76128 Karlsruhe, Germany
\vskip .2cm
\noindent Received   June 1997}

\parindent 0pt
M.-R. Li and P.J. Hirschfeld are grateful for support from the A.v. Humboldt
Foundation.


\begin{references}
\vspace*{ -1.7cm}
\bibitem{SD} N. Schopohl and O. V. Dolgov, Phys. Rev. Lett. 80, 4761(1998). 
\bibitem{impurities} F. Gross et al, Z. Phys. B64, 175 (1986).
\bibitem{PHW} W.O. Putikka, P.J. Hirschfeld, and P. W\"olfle, Phys. Rev. B
41, 7285 (1990).
\bibitem{KL} I. Kosztin and A.J. Leggett, Phys. Rev. Lett. 79, 135
(1997).
 \bibitem{Volovik} G.E. Volovik, cond/mat 9805159
\bibitem{Aminetal} M.H. S. Amin, I. Affleck, and M. Franz,
cond/mat/9712218
\bibitem{GK} D. I. Glazman and A. E. Koshelev, Sov. Phys. JETP 70, 774(1990).
 \end{references}
\end{document}